\documentclass[aps,prl,a4paper,reprint,showpacs,amsmath,amssymb,superscriptaddress,floats]{revtex4-1}

\usepackage{amssymb}
\usepackage{latexsym}

\usepackage{graphicx}

\usepackage{amsmath}
\usepackage{amsfonts}
\usepackage{bm}

\usepackage{color}

\newcommand{\be}{\begin{equation}}
\newcommand{\ee}{\end{equation}}
\newcommand{\bea}{\begin{eqnarray}}
\newcommand{\eea}{\end{eqnarray}}

\newcommand{\la}{\langle}
\newcommand{\ra}{\rangle}

\newcommand{\ri}{\mbox{i}}

\begin{document}
\title{Anderson Localization of Composite Excitations in Disordered Optomechanical Arrays}
\author{Thales Figueiredo Roque}
\affiliation{Instituto de F\'isica Gleb Wataghin, Universidade Estadual de Campinas, 13083-859 Campinas, S\~ao Paulo, Brazil}

\author{Vittorio Peano}
\affiliation{Institute for Theoretical Physics II, University Erlangen-N{\"u}rnberg, D-91058 Erlangen, Germany}

\author{Oleg M. Yevtushenko}
\affiliation{Institute for Theoretical Physics II, University Erlangen-N{\"u}rnberg, D-91058 Erlangen, Germany}

\author{Florian Marquardt}
\affiliation{Institute for Theoretical Physics II, University Erlangen-N{\"u}rnberg, D-91058 Erlangen, Germany}
\affiliation{Max Planck Institute for the Science of Light, G{\"u}nther-Scharowsky-Stra{\ss}e 1, D-91058 Erlangen, Germany}

\date{\today }

\begin{abstract}
      Optomechanical arrays are a promising future platform for studies of transport,
      many-body dynamics, quantum control and topological effects in systems of coupled
      photon and phonon modes. We introduce disordered optomechanical arrays, focusing on
      features of Anderson localization of hybrid photon-phonon excitations. It turns
      out that these represent a unique disordered system, where basic parameters
      can be easily controlled by varying the frequency and the amplitude of an external
      laser field. We show that the two-species setting leads to a non-trivial frequency
      dependence of the localization length for intermediate laser intensities. This could
      serve as a convincing evidence of localization in a non-equilibrium dissipative situation.
\end{abstract}


\maketitle

{\it Introduction}:
Optomechanics is a rapidly evolving research field at the intersection
of condensed matter and quantum optics \cite{OptoMechRev,Aspelmeyer2010}. By exploiting
radiation forces, light can be coupled to the mechanical motion of vibration modes.
The interplay of light and motion is now being used for a range of applications,
from sensitive measurements to quantum communication, while it also turns out to
be of significant interest for fundamental studies of quantum physics.

This rapidly developing area has so far mostly exploited the interaction between a
single optical mode and a single mechanical mode. Going beyond this, recent theoretical
research indicates the substantial promise of so-called optomechanical arrays, where
many modes are arranged in a periodic fashion. In such systems, a large variety of new
phenomena and applications is predicted to become accessible in the future. These include
the quantum many-body dynamics of photons and phonons \cite{Ludwig2013}, classical
synchronization and nonlinear pattern formation \cite{Heinrich2011,Holmes2012,Lauter2015},
tunable long-range coupling of phonon modes \cite{Xuereb2012,Xuereb2014,Schmidt2012},
photon-phonon polariton bandstructures and transport \cite{Clerk2014,Schmidt2015b},
artificial magnetic fields for photons \cite{Schmidt2015},
and topological transport of sound and light \cite{Peano2015}. A first experimental
realization of a larger-scale optomechanical array has recently been presented, involving
seven coupled optical microdisks \cite{Lipson2015}. Even greater potential is expected for
implementations based on optomechanical crystals \cite{Safavi-Naeini2010APL,Chan2012,SafaviNaeini2014},
i.e photonic crystals that can be patterned specifically to generate localized photon and phonon modes.

Given these promising predictions and the rapid experimental progress towards larger
arrays, the question of disorder effects now becomes of urgent importance. For example,
in the case of optomechanical crystals, experiments indicate fluctuations in the geometry
of about 1\%, which translate into equally large relative fluctuations of both the
mechanical and optical resonance frequencies. This will invariably have a very significant
impact on the transport properties. However, gaining a better understanding of disorder
effects in the various envisaged applications is only one motivation of the research to
be presented here. Of equal, possibly even greater, importance is the opportunity that
is offered by optomechanics to create a highly tuneable novel platform for deliberately
studying fundamental physical concepts such as Anderson localization \cite{AndLoc}.

Localization of waves in a random potential
is one of the most remarkable and nontrivial interference effects.
Initially, it has been studied in electronic disordered systems \cite{AndLocCM},
though this effect applies equally to other types of quantum and even classical waves
\cite{WavesRandMed}. By now, localization and related phenomena have been discovered
and investigated in photonic systems
\cite{AndLocPhot-0,AndLocPhot-1,AndLocPhot-2,AndLocPhot-3,AndLocPhot-4,AndLocPhot-5,RandLasAndLoc},
coupled resonator optical waveguides \cite{AndLocCROW}, cold atomic gases
\cite{AndLocGases-1,AndLocGases-2}, in propagation of acoustic waves \cite{AndLocAcWaves}
and in Josephson junction chains \cite{AndLocJosephson}.
Localization can even play a constructive role, namely in
random lasing \cite{RandLas,RandLasAndLoc}. In spite of extensive theoretical efforts, the unambiguous interpretation
of experimental manifestations of localization often remains a challenge, even in situations where the ideal version of Anderson localization applies.

Optomechanical arrays enable controlled optical excitation and readout and at the same
time promise significant flexibility in their design. However, it is the optical
tuneability of the interaction between two different species (photons and phonons) that
makes optomechanical systems a unique platform. As we will show in the present Letter,
this offers an opportunity to study effects in Anderson localization physics which currently
represent a significant challenge even on the theoretical level and will thus open the
door towards exploring novel physics that has not been observed so far.

\begin{figure}[t]
   \includegraphics[width=0.4 \textwidth]{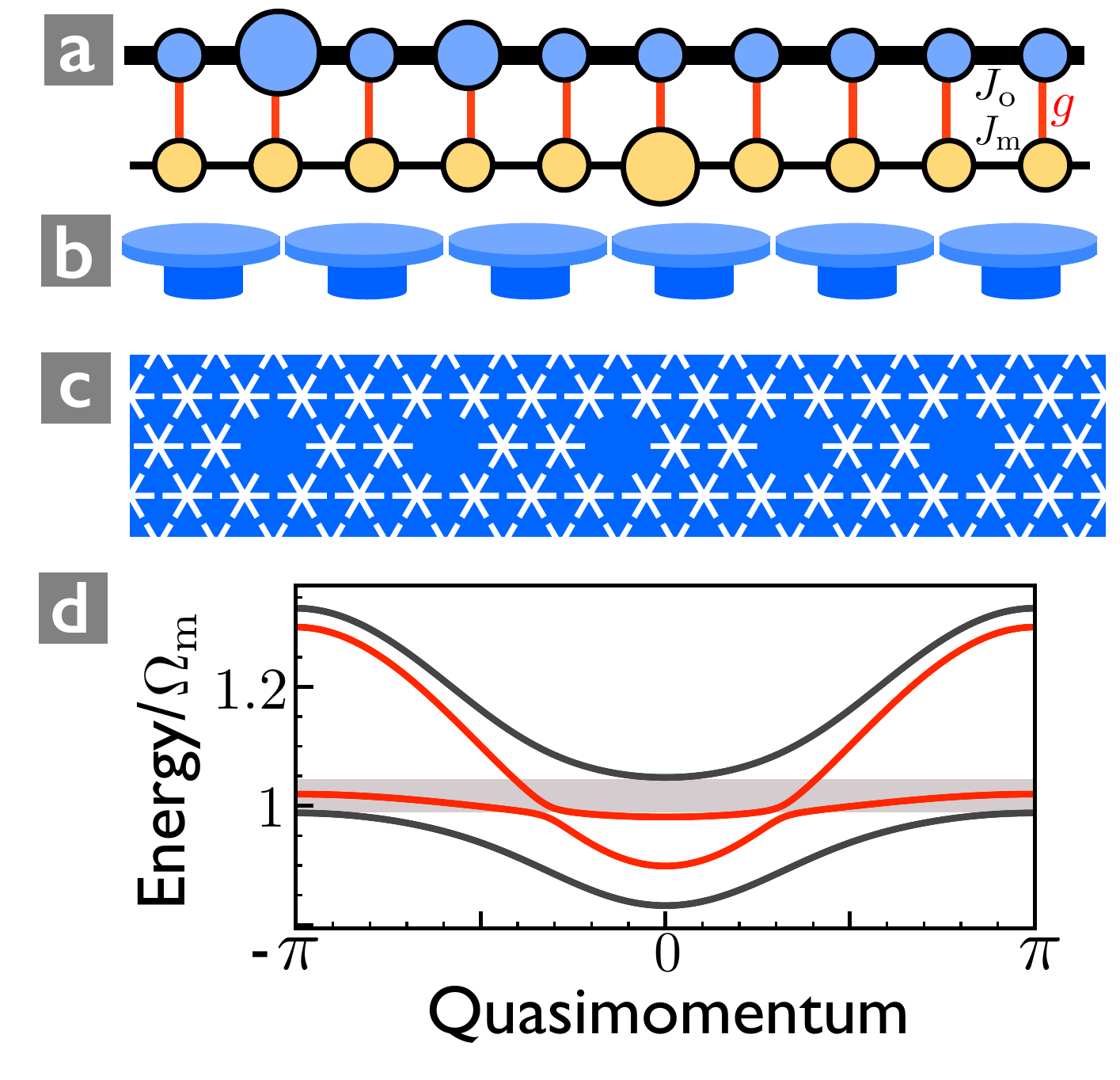}
   \caption{\label{OMA}
        (color on-line)
        Scheme, implementations and band structure of an optomechanical array.
        (a) An array of 
        photon (blue dots) and phonon (yellow) modes 
        can be viewed as a ladder:
        photons and phonons either can hop to nearest neighbor sites or
        can be interconverted on the same site. This system
        can be implemented by (b) an array of microdisks, or (c) an array of
        co-localized optical and mechanical defect modes in an optomechanical
        crystal. (d) Optomechanical band structure for two different values of the coupling
        $ g / \Omega_m = 0.01, 0.1 $ (red/black lines).
        For the larger interaction strength the upper and lower
        polariton bands are separated by a complete band gap (grey region).
        The other parameters are $ \Omega_o = 1.1 \Omega_m, J_o = 0.1
        \Omega_m, J_m = 0.01 \Omega_m $.
                 }
\end{figure}

{\it The model}:
We consider a 1D array of optomechanical cells (OMA), see Fig.\ref{OMA}, driven by
a single bright laser. The cell $j$ contains an optical and a vibrational mode that are coupled via the standard linearized
optomechanical Hamiltonian
\be
\hat{H}_j=\!\! \sum_{\nu=o,m} \!\! \omega_{\nu,j} \, \hat{n}_{\nu,j}  -
                                            g_j \left( \hat{c}_{o,j} + \hat{c}^\dagger_{o,j} \right) \left( \hat{c}_{m,j} +  \hat{c}^\dagger_{m,j} \right) ;
\label{eq:cellHamiltonian}
\ee
see Refs.\cite{OptoMechRev,OMA-model} for details.
Here $ \hat{n}_{\nu,j} \equiv \hat{c}^{\dagger}_{\nu,j} \hat{c}_{\nu,j} $, and
$ \hat{c}_{\nu,j}$ is the bosonic annihilation operator of either optical, $ \nu = ``o"$,
or mechanical, $\nu =  ``m"$, excitations (we set $\hbar =1$).
Due to disorder, the frequencies $ \omega_{\nu,j} $ fluctuate around mean values $ \la \omega_{\nu,j}
\ra_d = \Omega_\nu$. We assume that $ \omega_{\nu,j} $ are independent
Gaussian random variables with variances $ \la (\omega_{\nu,j} - \Omega_\nu) ( \omega_{\nu',j'} -
\Omega_{\nu'} ) \ra_d = \sigma_{\nu}^2 \delta_{j,j'} \delta_{\nu,\nu'} $.
Eq. (\ref{eq:cellHamiltonian}) is defined in a  rotating frame, where
the optical frequencies $\omega_{o,j}$ are counted off from the laser frequency, $ \omega_L $ \cite{OptoMechRev}.
Thus, $\Omega_o$ indicates the average detuning  and can be tuned in situ by
varying the laser frequency. The optomechanical couplings $g_j$ are proportional to the  mean amplitude of the
light circulating  in the cavity $j$ \cite{OptoMechRev}. Hence, they are also tunable by varying
the laser power.


The presence of two-mode squeezing interactions in Eq. (\ref{eq:cellHamiltonian}) can in principle lead to instabilities.
We choose $\Omega_o$ such that these terms are off-resonant and disorder configurations
with optical \cite{Q-Opt} or vibrational instabilities are very rare. We leave their study for a
forthcoming paper.

We can describe the full OMA by a Hamiltonian with nearest-neighbor optical, $J_{o}$, and
mechanical, $J_{m}$, hopping amplitudes:
\be
\label{Hom}
  \hat{H}= \sum_j\hat{H}_j - \hat{H}_{\rm h} , \quad
  \hat{H}_{\rm h} = \sum_{j,\nu}  J_{\nu}  \hat{c}^{\dagger}_{\nu,j+1} \hat{c}_{\nu,j} + H.c.
\ee
Our model is time-reversal symmetric \cite{TRS}.



{\it Clean polariton bands}:
In a clean
OMA without dissipation (and without squeezing interaction),
the  photon--phonon hybridization leads to a pair of bands with energies
 \be
  \Omega_\pm =\bar{\Omega} -2\bar{J} \cos(k) \pm
                   \sqrt{ \left[\delta \Omega/2 - \delta J \, \cos(k)   \right]^2 + g^2} ,
\ee
where $\bar{\Omega}=(\Omega_o+\Omega_m)/2$, $\delta\Omega=(\Omega_o-\Omega_m)$ and
likewise for $\bar{J}, \, \delta J$. We refer to $\Omega_\pm$ as upper/lower polariton band, respectively.
$ k $ denotes the wave-vector of polaritonic Bloch states.
We focus on the regime where the uncoupled bands overlap, $ \delta\Omega < 4 \bar{J}$.
In this case, the polariton bands are separated by a gap if the coupling becomes large enough,
$ g > g_{\rm min} $ \cite{Gmin}, see Fig.\ref{OMA}.

{\it Anderson localization of uncoupled excitations}:
It is well known that even weak disorder leads to a crucial effect in a
1D system: the eigenstates become localized. If $ g = 0 $,
each subsystem (photon/phonon), is individually described by the 1D Anderson model \cite{AndLoc}.
The localized states
decay exponentially away from their center, $ \sim \exp(-|j-j_0|/\xi^{(0)}_{\nu}) $. Here $ \xi^{(0)}_{o,m} $ are the bare localization
lengths for photons and phonons (for $g=0$), measured in units of the lattice constant. Using the theory of 1D localization
\cite{1D-loc}, we can approximate the frequency dependence of the localization length:

\be
\label{BareXi}
 \xi^{(0)}_{\nu}(\Omega) \simeq 2 ( 2 \sin[k_\nu(\Omega)] / \chi_\nu )^2 ;
\ee
here the dimensionless quantities $ \chi_\nu \equiv \sigma_\nu / J_\nu $ and $ 2 \sin[k_{\nu}] $ are
the disorder strength and the bare group velocity, respectively.
Eq.(\ref{BareXi}) is valid for weak (up to moderately strong) disorder \cite{WeakDis}. The comparison
of Eq.(\ref{BareXi}) with numerical results is shown below in Fig.\ref{LocLenEn}.

In any experiment, localization
can be detected if photons and phonons explore the localization length before leaking out, at a rate $\kappa_{o,m}$. This holds true if
$ \xi_\nu < 2 J_\nu | \sin(k_\nu) | / \kappa_\nu $,
which allows us to neglect dissipation in a first approximation \cite{BallProp}.
In addition, the sample size $L$ should be larger than the localization length, $ L \gg {\rm max}(\xi^{(0)}_\nu) $. For typical $ L \sim 100 $ we need $ {\rm max}(\xi^{(0)}_\nu) \sim 10 $, corresponding to $ \chi_\nu \sim 1 $.


\begin{figure}[t]
   \includegraphics[width=0.375 \textwidth]{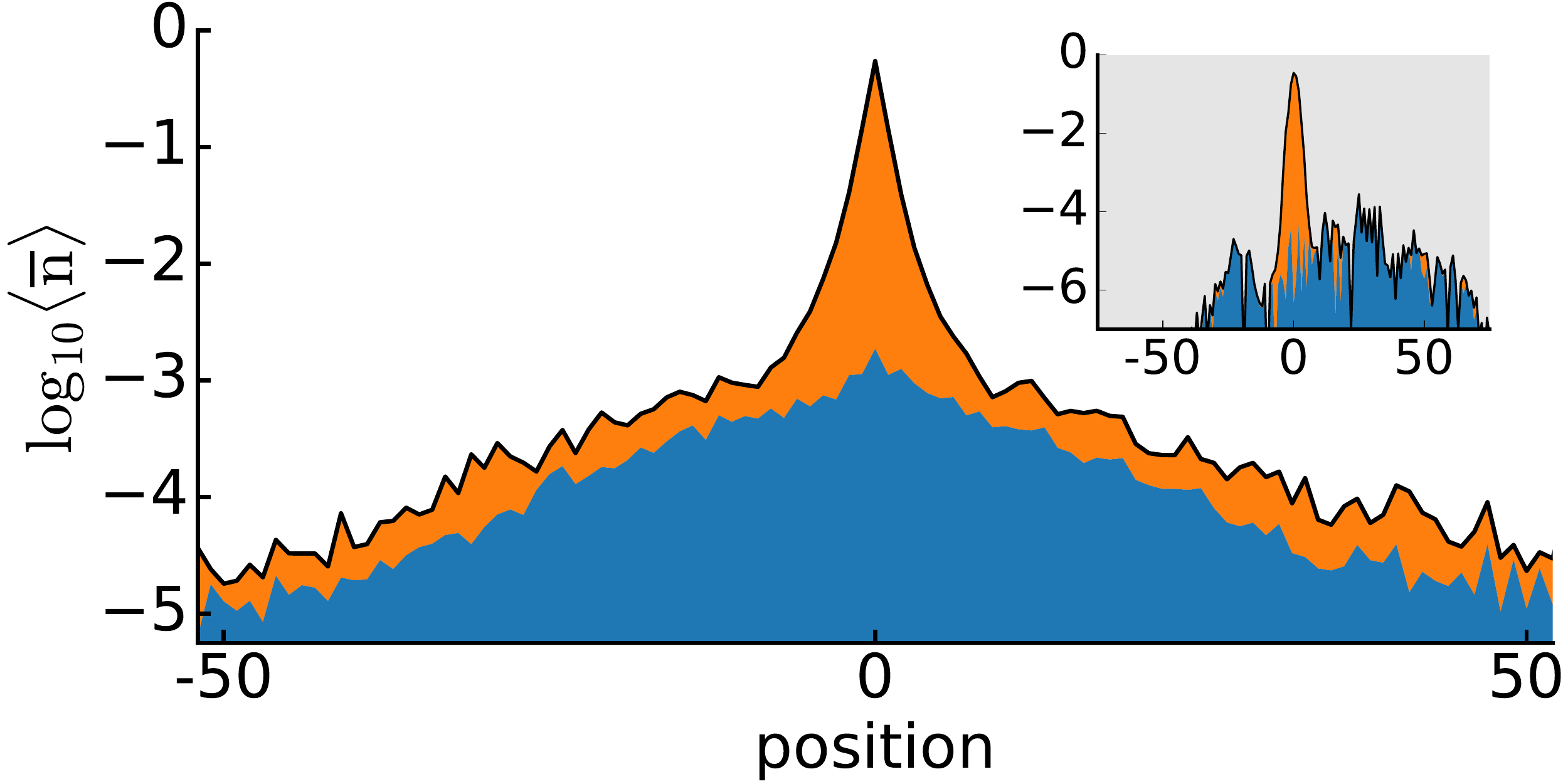}
   \includegraphics[width=0.375 \textwidth]{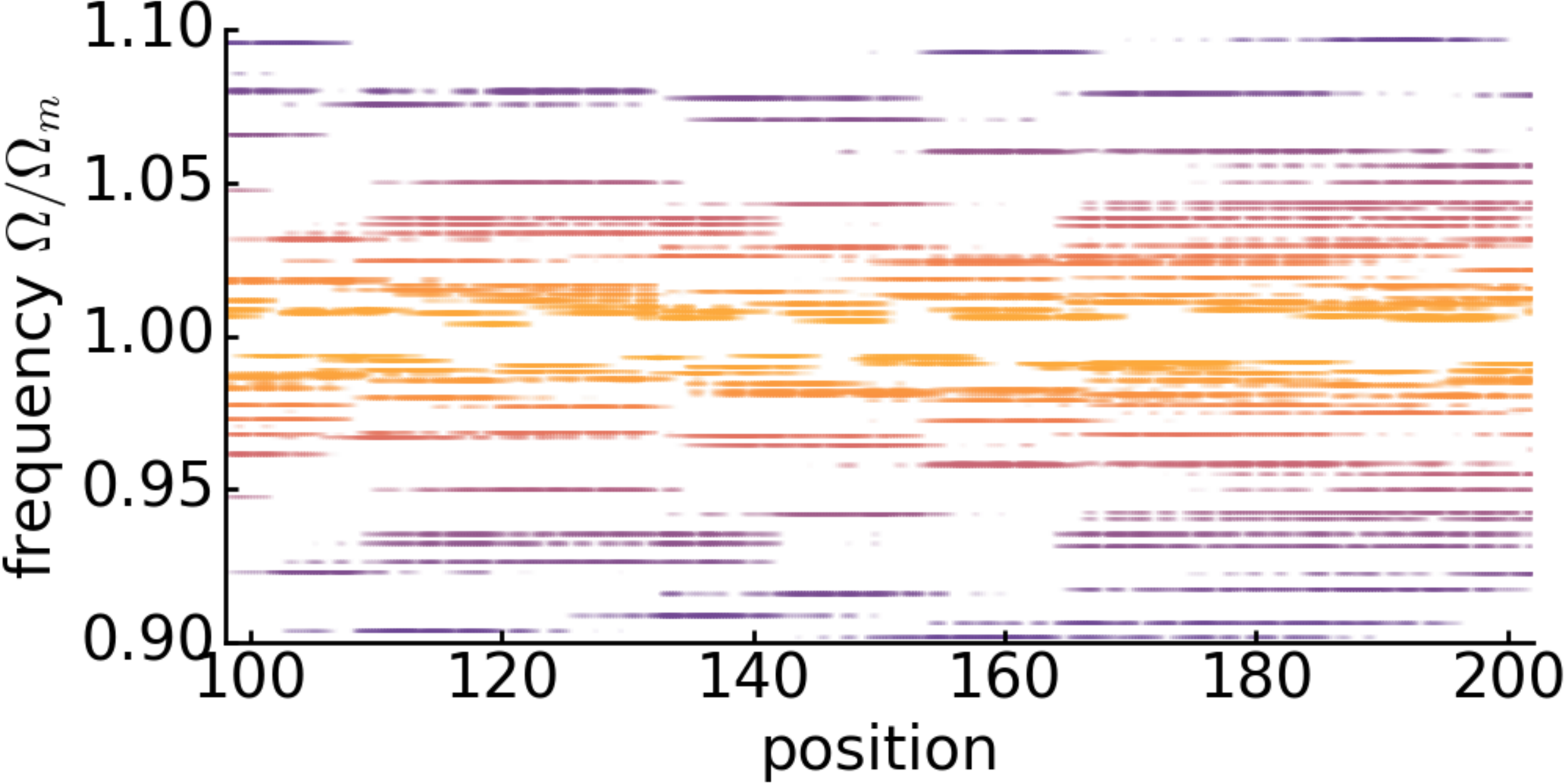}
   \caption{\label{WFunc}
        (color on-line)
        Upper panel: Typical shape of an eigenstate,
        at $ \Omega_o = \Omega_m, \ g = 0.001 \Omega_m $, without disorder averaging (inset) and after disorder
        averaging over 500 realizations (main picture). Other parameters are explained in the text. Two different localization lengths
        are clearly visible. The profiles of $ \bar{n}_j $, obtained from different disorder realizations,
	  have been shifted in space, such that their maxima are always located at $ j = 0 $.
        Lower panel: Local density of states (frequency- and position-resolved spectrum) at $ g = 0.05 \Omega_m $; other parameters
        as in the upper panel. The color ranges from orange to blue, depending
        on whether a given eigenstate has a stronger mechanical or optical component, respectively.
                 }
\end{figure}

{\it Localization in Optomechanical Arrays}:
At finite pho\-ton-phonon coupling, we encounter an Anderson model with two channels. Localization in the symmetric
version of this model (with equal
parameters of each channel) is well studied and understood
\cite{Dorokhov,MPK}. However, OMAs do not fall into this universality class since the mechanical
band is generically much narrower than the optical one, $ J_m \ll J_o $. Thus, the hybrid
excitations consist of two components with very different velocities.
Similar composite quasiparticles are not uncommon, another example is given by cavity polaritons
\cite{Polaritons-1,Polaritons-2} including polaritons in a disordered potential \cite{Polaritons-3}.
Developing the theory of localization for such non-symmetric systems remains a real challenge,
cf. Ref.\cite{KrLadders}.
The hybrid localized states typically have two localization lengths, $ \xi_1 < \xi_2 $,
see the upper panel of Fig.\ref{WFunc}.
For small systems, $ L < \xi_1 $, the excitations do not feel localization
and propagate ballistically. Their transmission decays as $ \exp(-L/\xi_1) $
in the range $ \xi_1 < L < \xi_2 $ and becomes suppressed as $ \exp(-L/\xi_2) $ at $ L > \xi_2 $. Our numerical analysis
shows that the space region where $ \xi_1 $ dominates quickly shrinks with increasing $ g $.
Therefore, $ \xi_2 $ seems more interesting experimentally, and we will focus on this 'large' localization length in the following.
We start from a numerical analysis for relatively strong disorder. At the
first stage, we neglect disorder-induced fluctuations of $ g_j $ \cite{CouplingFluct} and use its homogeneous
mean value $ g = {\rm const} $.

{\it The method}:
The localization length can be obtained, e.g., from  the photon-photon transmission,
$T_{oo}(j,k;\Omega) \propto\big| G^R_{oo}(j,k;\Omega) \big|^{2}$ where
$G^R_{oo}(j,k;\Omega)= $ $ -i\int_{0}^\infty dt\exp (i \Omega t) \overline{ [\hat{c}_{o,j}(t), \hat{c}_{o,k}^\dagger(0)] }$
is the frequency-resol\-ved retarded Green's function. $ T_{oo} $ is defined via the optical power
detected on site $ j $ at frequency $\omega_L+ \Omega$ while a probe laser of the same frequency is impinging
on a different site $ k $ \cite{LocLen-Calc}. For $ x=|j-k| \to \infty $, we expect $ T_{oo}(j,k;\Omega) \propto
\exp(- 2 x/\xi_2) $. Thus, the expression for the averaged (inverse) localization length reads
\be
\label{LocLen}
  \xi^{-1}_2(\Omega)= -\lim\limits_{x \to \infty} \left(
                           \left\la \ln\bigl(T_{oo}(j,k;\Omega) \bigr) \right\ra_d
                           \Bigl/ 2 x
                                              \right).
\ee
We note that the value of $ \xi_2(\Omega) $ is the same for other transmission
processes (e.g. photon-phonon transmission) \cite{LocLen-Calc}.

Eq.(\ref{LocLen}) can be used as a definition even in the presence of dissipation.
In the absence of dissipation and instabilities, there is a simpler alternative, namely extracting the localization
length directly from the spatial profile of eigenstates \cite{LocLen-Calc}.
To ensure reliability of results, we have combined both approaches in numerical simulations.


{\it Analysis of numerical results}:
The upper panel of Fig.\ref{WFunc} shows a typical optomechanical eigenstate
in the case of small coupling. The excitation frequency has been selected from the tail of the pure
mechanical band. Two different slopes, which correspond to two different localization lengths
$ \xi_{1,2} $, are clearly visible. When $ g $ increases and the other parameters of the upper Fig.\ref{WFunc}
remain unchanged,
the region where $ \xi_1 $ dominates shrinks \cite{MainLocRad} and becomes invisible
very quickly. In the following, we will concentrate on $ \xi_2 $
and will denote it as $\xi$ for the sake of brevity. The lower panel of Fig.\ref{WFunc} illustrates
the distribution of optomechanical excitations in space and frequency, including the character of excitations
(photon vs. phonon).

Here, and in the following, we have displayed numerical results for an illustrative
set of parameters: $ J_o = 0.1 \Omega_m \gg \ J_m = 10^{-3}
\Omega_m $. Localization of the optomechanical excitations becomes pronounced
at $ \chi_{o,m} \sim 1 $. For concreteness, we have chosen
equal relative disorder strength, $ \chi_o = \chi_m = 1 $. In real samples, $ J_o $ ranges 
from 1GHz to 10THz ($ J_m $: from 100kHz to 1GHz) with the optical disorder being of 
order 100GHz to 1THz  (mechanical: from 10MHz to 100MHz). Thus, our choice of 
$ \chi_{o,m} $ falls into the range of experimentally relevant parameters.
The optomechanical coupling in our numerics ranges from weak,
$ g = 10^{-3} \Omega_m $, to strong, $ g = 0.05 \Omega_m $. To suppress finite size effects,
we employed large systems, $L=10^3 \gg \xi $, during exact diagonalization. The Green's
functions method has allowed us to explore even much larger sizes.


In Fig.\ref{LocLenEn}, we display the frequency-dependence of the localization length of hybrid optomechanical
excitations in a disordered array, one of the central numerical results of this article. For comparison, we also show the
situation for the uncoupled systems, including the (scaled) analytical expression for $ \xi^{(0)}_{o} $,
Eq.(\ref{BareXi}) \cite{Scaling} (green solid line in Fig.\ref{LocLenEn}a).
Once the subsystems are coupled, significant changes of $ \xi(\Omega) $ occur in the vicinity of the
unperturbed (narrow) mechanical band where the optomechanical hybridization is most efficient.
Firstly we note that, if $ 0 < g < \Delta_{\rm loc}^{(m)} $, the coupling between the optical and the mechanical 
systems is perturbatively weak even in the middle of the mechanical band [region I in Fig.\ref{LocLenEn}(f)].
On the other hand, when the optomechanical coupling becomes large,
$ g > \sqrt{J_o J_m} = \sqrt{\sigma_o \sigma_m} \sim  \Delta_{\rm loc}^{(o)} $ for our choice of parameters,
a gap opens around the resonant frequency $ \Omega = \Omega_m $ and remaining excitations inside the gap tend to 
become localized [Fig.\ref{LocLenEn}(e)].

\begin{widetext}

\begin{figure}[ht]
\includegraphics[width=0.975 \textwidth]{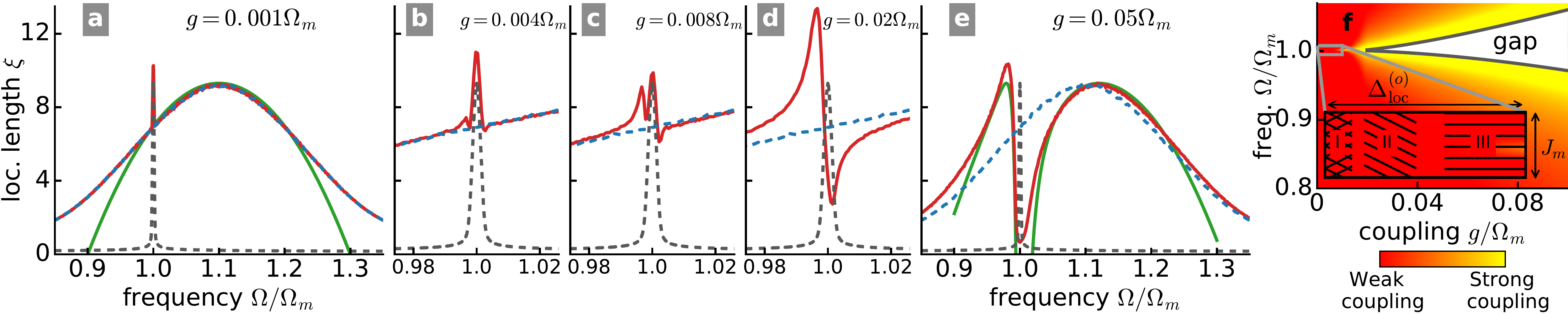}
   \caption{\label{LocLenEn}
        (color on-line)
        Frequency-dependence of the localization length: Dashed lines show bare ($ g = 0 $)
        optical ($ \xi_o^{(0)} $ - blue)
        and mechanical ($ \xi_m^{(0)} $ - black) localization lengths. The red solid line shows
        the localization length of hybrid excitations, $ \xi $, calculated at several values of the
        optomechanical coupling, $ 0.001 \Omega_m \le g \le 0.05 \Omega_m $, and $ \Omega_o = 1.1
        \Omega_m $. Green solid lines describing $ \xi^{(0)}_o $ in panel (a) and $ \xi $ in panel (e) are
        obtained from Eqs.(\ref{BareXi},\ref{EnDep}), respectively, after scaling by a constant factor.
       Panel (f) illustrates schematically the different regimes as a function of coupling and frequency.
%
                 }
\end{figure}

\end{widetext}

Analytical methods which would allow one to explore localization in strongly disordered systems are not available in general.
Nevertheless, it turns out that our optomechanical array corresponds to a certain two-channel system,  which was studied
analytically in Ref.\cite{KrLadders} for the limit of weak disorder and large coupling. Remarkably, the shape of our numerically
extracted $ \xi(\Omega) $ at large $ g $ agrees with the predictions of Ref.\cite{KrLadders}, even though we are here dealing with
strong disorder, $ \chi_{\nu} \sim 1 $ \cite{MPK}.
The theory of Ref.\cite{KrLadders} is valid if $ g $ is large compared with the (bare) mean level
spacing in the localization volume, $ \Delta_{\rm loc}^{(\nu)} $, which holds true for the parameters 
of our numerical study at $ g \ge 0.05 \Omega_m $  \cite{Validity-g}. If $ g > g_{\rm min} $,
(i.e., if the clean polariton bands are separated by the gap of the width $ \Omega_+(k=0) - \Omega_-(k=\pi) $)
we can use the following (leading in $ \chi_\nu $) expression for the localization length \cite{KrLadders}:
\bea
 \label{EnDep}
   \xi( \Omega ) & \simeq & 4 \bigl(2 \sin \left[ k_\pm(\Omega) \right] \bigr)^2 \bigl/
                                                   \bigl( \chi^2 \bigl[ 1 + \cos^2(\gamma) \bigr] \bigr) \, ; \\
   \tan(\gamma) & = & 2 \sqrt{J_o J_m} g / \delta J (\Omega - \Omega_r) , \quad
                        \Omega_r \equiv J_o \delta\Omega / \delta J .
   \nonumber
\eea
Here $ \chi =  \chi_o / {\cal C} = \chi_m /  {\cal C} $ and
$ k_\pm(\Omega) $ denotes the inverted dispersion relation  $ \Omega_\pm(k) $. The quantity
$ V_\pm \equiv 2 \sin \left[ k_\pm(\Omega) \right] $ is called ``rapidity''. It coincides
with the group velocity of the excitations for $ g = 0 $,
and according to Eq.(\ref{EnDep}) it governs the frequency-dependence of $ \xi( \Omega ) $ in the coupled case.
The factor $ {\cal C} $ reflects renormalization of the disorder strength caused by the optomechanical
coupling. Calculation of $ {\cal C} $ is beyond the scope of Ref.\cite{KrLadders}
and we have found its approximate value $ {\cal C} \simeq 1.16 $ by fitting the analytically
calculated maximal value of $ \xi(\Omega>\Omega_m) $ to the numerical one.
Fig.\ref{LocLenEn}e shows the comparison of the analytical and numerical results. They differ
noticeably only close to edges of the clean band where the analytical theory
looses its validity because $ \xi \to 1 $. In addition, the gap is smeared by the relatively strong disorder.

We have discovered that, at  $ \Omega\simeq\Omega_m $,  the crossover between small and large 
values of $ g $ is highly non-trivial (and it is outside the scope of the analytical theory): 
when the optomechanical coupling increases from $ g \sim \Delta_{\rm loc}^{(m)} $ to $ g \sim J_m $
[region II in Fig.\ref{LocLenEn}(f)], the single maximum of $ \xi $ [cf. Fig.\ref{LocLenEn}(a)$ \to $(b)]
grows sublinearly in $ g $ \cite{Small-g}.
This growth  stops and turns into a decrease when $ g \gg J_m  $. Simultaneously, a new local maximum 
develops at the frequency corresponding to the maximum of the rapidity  [Fig.\ref{LocLenEn}(b)$ \to $(c)
and region III in Fig.\ref{LocLenEn}(f)]. Finally, the new local maximum becomes the global one and a dip 
appears close to $ \Omega_m $ at $ g \ge \Delta^{(o)}_{\rm loc} $
[Fig.\ref{LocLenEn}(c)$ \to $(d)]. This non-trivial dependence of the localization length on the
coupling constant, i.e., on the tuneable intensity of the external laser, could help to distinguish localization
and trivial dissipation effects in real experiments.

\begin{figure}[t]
   \includegraphics[width=0.47 \textwidth]{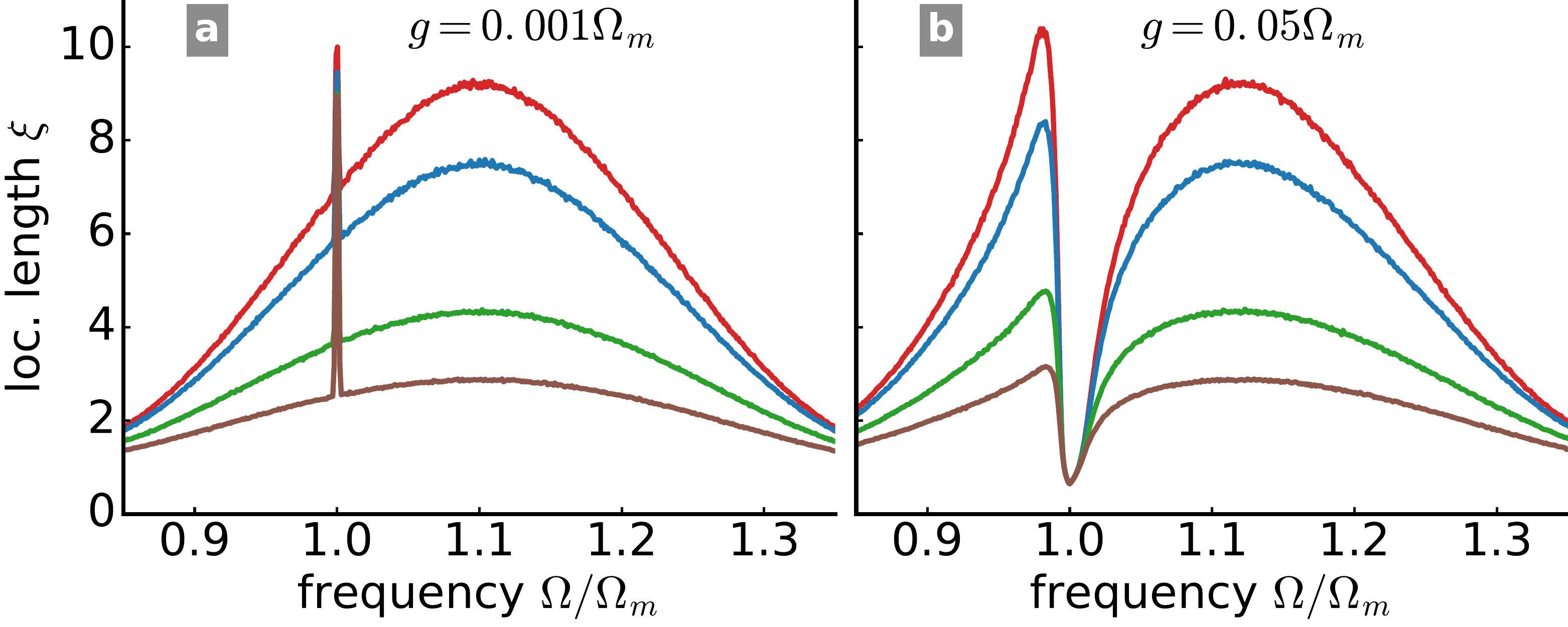}
   \caption{\label{Dissip}
        (color on-line)
        Frequency-dependence of the localization length at $ \Omega_o = 1.1 \Omega_m $
        and $ g / \Omega_m = \{ 0.001, 0.05 \} $ 
        calculated for different values of the optical decay rate
        $ \kappa_o / \Omega_m = \{ 0, 0.01, 0.05, 0.1 \} $ (red-, blue-, green-, and brown lines,
        respectively). Note that the peak at $ g = 0.001 \Omega_m $ and $ \Omega = \Omega_m $
        is almost insensitive to the optical dissipation since the corresponding wavefunctions
        have mainly mechanical components.
           }
\end{figure}

We have checked that the shape of $ \xi( \Omega ) $ is robust with respect to dissipation effects
as long as the mean level spacing in the localization volume of the hybrid excitations is larger
than the optical and mechanical decay rates, $ \kappa_\nu $ \cite{DissipEsc}.
Propagation of the excitations is suppressed due to their
finite life time which is reflected by the frequency-independent decrease of $ \xi $. Typical profiles
$ \xi( \Omega ) $ are shown in Fig.\ref{Dissip} where the optical dissipation rate increases until
$ \kappa_o = 0.1 \Omega_m $. These profiles are also robust with respect to the spatial inhomogeneity
of $ g_j $ which results from randomness of the cell frequencies \cite{CouplingFluct,Fluct-g}.

{\it Conclusions and discussion}: Disordered OMAs belong to a new class of disordered systems
where composite (photon-phonon) excitations are localized and
the most important parameters can be easily fine-tuned.
Thus, OMAs provide a unique opportunity to study Anderson localization of composite
particles in real experiments. Moreover, they should allow to reliably distinguish localization from trivial dissipation effects.
Future studies may address the additional novel physics that will arise when two-mode
squeezing processes become relevant. At strong driving, this could involve the interplay between
instabilities and localization, with interesting connections to random lasing, extending the new research domain of disordered optomechanical arrays into the nonlinear regime.

\begin{acknowledgments}
We  acknowledge support from the EU Research Council through the grant
EU-ERC OPTOMECH 278320. TFR acknowledges support from
FAPESP. We are grateful to Vladimir Kravtsov and Igor Yurkevich
for useful discussions.
\end{acknowledgments}

\bibliography{Bibliography,Bibliography-OptMech}

\begin{thebibliography}{56}
\bibitem [{OMA()}]{OMA-model}%
  \BibitemOpen
  \href@noop {} {}\bibinfo {note} {Derivation of the standard optomechanical
  Hamiltonian is briefly reviewed in Suppl.Mat.1}\BibitemShut {NoStop}%
\bibitem [{TRS()}]{TRS}%
  \BibitemOpen
  \href@noop {} {}\bibinfo {note} {This means that $ J_{o,m}, g_0 $ and $ g $
  are real.}\BibitemShut {Stop}%
\bibitem [{Gmi()}]{Gmin}%
  \BibitemOpen
  \href@noop {} {}\bibinfo {note} {{The value of $ g_{\rm min} $ is found from
  the condition $ \Omega_+(k=0) = \Omega_-(k=\pi) \Rightarrow g_{\rm min} = 2
  {\rm Re} \sqrt{J_o J_m \left( 1 - [\delta \Omega / 4 \bar{J}]^2 \right) } $
  }}\BibitemShut {NoStop}%
\bibitem [{Wea()}]{WeakDis}%
  \BibitemOpen
  \href@noop {} {}\bibinfo {note} {{Strictly speaking, one must require $
  \chi_{\nu} < 1 $ though a softer condition $ \xi_\nu \gg 1 $ sufficies for
  practical purposes.}}\BibitemShut {Stop}%
\bibitem [{Bal()}]{BallProp}%
  \BibitemOpen
  \href@noop {} {}\bibinfo {note} {{We have taken into account that, in 1D
  systems, the excitations propagate ballistically in the localization
  volume.}}\BibitemShut {Stop}%
\bibitem [{Cou()}]{CouplingFluct}%
  \BibitemOpen
  \href@noop {} {}\bibinfo {note} {$ g_j $ is proportional to the mean
  occupation number of the photons on the site $ j $ while the cells with
  smaller optical frequencies host more photons. This locally enchances $ g_j $
  on these sites.}\BibitemShut {Stop}%
\bibitem [{Loc()}]{LocLen-Calc}%
  \BibitemOpen
  \href@noop {} {}\bibinfo {note} {Algebraic details for the derivation of the
  localization radius can be found in Suppl.Mat.2.}\BibitemShut {Stop}%
\bibitem [{Mai()}]{MainLocRad}%
  \BibitemOpen
  \href@noop {} {}\bibinfo {note} {{If $ g $ increases and the other parameters
  of the upper Fig.\ref{WFunc} remain unchanged, $ \xi_2 $ slightly increases.
  This might lead to suppression of the region where $ \xi_1 $ dominates
  because the eigenstate is normalized.}}\BibitemShut {Stop}%
\bibitem [{Sca()}]{Scaling}%
  \BibitemOpen
  \href@noop {} {}\bibinfo {note} {{The accuracy of Eq.(\ref{BareXi}) is
  insufficient to reproduce the maximal value of $ \xi^{(0)}_o(\Omega_o) $ at
  the relatively strong optical disorder, $ \chi_o = 1 $. Therefore, we have
  scaled the analytical answer by a constant factor to adjust the
  heights.}}\BibitemShut {Stop}%
\bibitem [{Val()}]{Validity-g}%
  \BibitemOpen
  \href@noop {} {}\bibinfo {note} {The mean level spacing in the localization
  volume is obtained from the relation $ \Delta_{\rm loc} / \Delta = N / \xi $
  . Using the estimate for the mean level spacing in a band of extremely
  localized states $ \Delta \sim \delta/N $ \cite{YeKr-virial,YeKr-DoS}, we
  find $ \Delta^{(\nu)}_{\rm loc} \sim \delta_{\nu} / \xi^{(0)}_{\nu} \sim
  J_\nu / \xi^{(0)}_{\nu} $. Here $ \xi^{(0)}_{\nu} $ denotes bare (at $ g = 0
  $) localization lengths. For the chosen parameters, this yields the condition
  $ g > 10^{-1} {\rm max} \{ J_\nu \} \sim 10^{-2} \Omega_m $. Thus, the
  analytical theory from Ref.\cite{KrLadders} can be used to understand the
  case $ g = 0.05 \Omega_m $ while smaller values of $ g $ are beyond the
  validity range of the analytic expressions.}\BibitemShut {Stop}%
\bibitem [{Sma()}]{Small-g}%
  \BibitemOpen
  \href@noop {} {}\bibinfo {note} {{The increase of $
  \xi_2(\Omega\simeq\Omega_m) $ when the optomechanical coupling is small, $ g
  < J_m $, is explained in Suppl.Mat.3.}}\BibitemShut {Stop}%
\bibitem [{Dis()}]{DissipEsc}%
  \BibitemOpen
  \href@noop {} {}\bibinfo {note} {{Equivalently, we can require that the
  localization length is much smaller than the escape length.}}\BibitemShut
  {Stop}%
\bibitem [{Flu()}]{Fluct-g}%
  \BibitemOpen
  \href@noop {} {}\bibinfo {note} {{This robustness is exemplified in
  Suppl.Mat.4.}}\BibitemShut {Stop}%
\end{thebibliography}


\onecolumngrid

\protect\pagebreak

\section*{Supplemental Materials}

\subsection*{1. Standard optomechanical Hamiltonian}

The linearized Hamiltonian, Eq.(\ref{Hom}) of the main text, is derived starting from the  Hamiltonians
of  a phononic and a photonic array. In the tight-binding approximation, both  Hamiltonians take the same form,
\be\label{eq:isarray}
  \hat{H}_{\nu} = \sum_j \left\{ \omega_{\nu,j} \, \hat{c}^{\dagger}_{\nu,j} \hat{c}_{\nu,j} -
                                                 J_{\nu} \left( \hat{c}^{\dagger}_{\nu,j+1} \hat{c}_{\nu,j} + H.c.
                                       \right) \right\}.
\ee
Here, the index $\nu =  ``o"$ and $\nu =  ``m"$ refers to the optical and mechanical  degrees  of freedom. The operators
 $ \hat{c}_{\nu,j} $ denote the  annihilation operator for the site $ j $; $ \omega_{\nu,j} $ and  $ J_\nu $
denote random on-site frequencies and constant overlap integrals, respectively.
The optical and mechanical modes co-localized on the same site are coupled by the  radiation pressure  force. The resulting interaction reads
\be
\label{Coupl}
   \hat{H}_{om} = - g_0 \sum_j   \hat{c}^{\dagger}_{o,j} \hat{c}_{o,j} (\hat{c}^{\dagger}_{m,j} + \hat{c}_{m,j}) \, ,
\ee
where $g_0$ is the   eigenfrequency shift of a localized  optical mode  by   a single phonon on the same site.
The presence of  a laser drive  of frequency $\omega_L$ is described by the additional Hamiltonian term
\be
   \hat{H}_{\rm laser} = \alpha_L \sum_j \hat{c}_{o,j} \exp (i \omega_L t) + H.c.
\ee
 The dynamics of the OMA with the Hamiltonian $ \hat{H} = \hat{H}_{o} + \hat{H}_{m} + \hat{H}_{\rm om}
+ \hat{H}_{\rm laser} $ is most conveniently described in the rotating frame defined by the unitary transformation
\be
  \hat{H} \to \hat{U} \hat{H} \hat{U}^\dagger - i \, \hat{U}  \frac{d}{dt}\hat{U}^\dagger \, , \quad
  \hat{U} =\sum_j \exp\left( i \omega_L t \, \hat{c}^\dagger_{0,j} \hat{c}_{0,j} \right) .
\ee
We decompose the operators $\hat{c}_{\nu,j}$ as sums of their mean filed values and new displaced
operators, $\delta \hat{c}_{\nu,j}$, incorporating the fluctuations, $\hat{c}_{\nu,j}=\langle \hat{c}_{\nu,j}\rangle+\delta \hat{c}_{\nu,j}$.
After inserting this decomposition into $ \hat{H} $,
all linear terms in $\delta\hat{c}_{\nu,j}$ cancel out in the Hamiltonian. We, thus, reproduce Eq.(\ref{eq:isarray}) but
with the fluctuation operators $\delta\hat{c}_{\nu,j}$ and the detunings $\tilde{\omega}_{o,j}-\omega_L$ replacing
the bare operators $\hat{c}_{\nu,j}$ and the optical frequencies  $\omega_{o,j}$, respectively. The  eigenfrequencies $\tilde{\omega}_{o,j}$ of the optical localized modes   include a small power-dependent frequency shift due to a static displacement $\propto 2{\rm Re}[\langle\hat{c}_{m,j}\rangle]$ of the corresponding mechanical oscillators, $\tilde{\omega}_{o,j}=\omega_{o,j}-2g_0{\rm Re}[\langle\hat{c}_{m,j} \rangle]$. In the limit where the
fluctuations $\delta\hat{c}_{\nu,j}$ are small compared to the mean values (for a strong enough drive), we can neglect
all cubic terms
and arrive to the linearized opto-mechanical interaction \cite{OptoMechRev}
\be
\label{Hmfa}
   \hat{H}_{\rm om}^{\rm(L)} = - \sum_j  \left(g^*_j \delta \hat{c}_{o,j} + H.c. \right)
                                                                             \left( \delta\hat{c}_{m,j} + H.c. \right) \, .
\ee
Here $ \, g_j \equiv g_0 \, \langle \hat{c}_{o,j}\rangle$ are the couplings of the linearized interaction.  In the main text, we have investigated a parameter regime where the laser is red-detuned compared to all optical resonances. We  have also focused on the OMAs where the broadening of the resonances (set by the typical decay rate $ \kappa_o $) is smaller than the minimal detuning. In this case,  all $g_j$ are real valued, consequently,
 the time-reversal symmetry is preserved by the OM interaction.
Summing all contributions, we obtain the Hamiltonian $ \hat{H} $ of the main text. There, for brevity,
we use $\hat{c}_{\nu,j}$ and $\omega_{o,j}$ for the fluctuation operators and the detunings, respectively.
We follow this convention also below. In the main text, we have also assumed that all linearized couplings $g_j$ are approximately equal, $g_j\approx g$.  This approximation holds when the mean value of the onsite detuning is much larger than its typical fluctuations, the optical hopping rate, and the typical  optical decay rate.  Below, we go beyond this approach investigating fluctuating coupling constants $g_j$.

\subsection*{2. Calculation of the localization length}

\subsubsection*{2.1 Input/Output formalism}

Let us express the elastic part of the photon-photon transmission $T_{oo}(k,j;\Omega) $ in terms of the retarded
(photon-photon) Green's function.

The response of the OMA to an additional probe field is described by the standard Langevin
equations [S1]:
\begin{equation}\label{eq:langevin}
 \dot{\hat{c}}_{\nu,j}=i[\hat{H},\hat{c}_{\nu,j}]-\kappa_{\nu, j}\hat{c}_{\nu,j}/2+\sqrt{\kappa_{\nu, j}}\hat{c}^{\rm in}_{\nu,j};
\end{equation}
where $\hat{c}^{\rm in}_{\nu,j}$ is the input (probe) field. The corresponding output field $\hat{c}^{\rm out}_{\nu,j}$
is given by the input/output relations
\begin{equation}\label{eq:input/output}
 \hat{c}^{\rm out}_{\nu,j}=\hat{c}^{\rm in}_{\nu,j}-\sqrt{\kappa_{\nu, j}}\hat{c}_{\nu,j}.
\end{equation}
For  a probe laser of frequency $\omega_L+\Omega$ (corresponding to the frequency $\Omega$ in the rotating frame) applied at site $ k $ we have
\begin{equation}\label{eq:cinput}
   \overline{ \hat{c}^{\rm in}_{o,l} } = \delta_{l,k}\alpha_p e^{-i \Omega t}, \quad
   \overline{ \hat{c}^{\rm in}_{m,l} } = 0,
\end{equation}
where $ \alpha_p $ is the amplitude of the probe laser.

The transmission is defined with the help of the ratio
\be
\label{NormAmp}
   \overline{ \hat{c}^{\rm out}_{o,j} } \Bigl/ \overline{ \hat{c}^{\rm in}_{o,k} } =
        \delta_{j,k} - \sqrt{\kappa_{o,j}} \ \overline{ \hat{c}_{o,j} } \Bigl/ \overline{ \hat{c}^{\rm in}_{o,k} }.
\ee
The Hamiltonian has been linearized, hence, the response of $ \overline{ \hat{c}_{o,j} } $ to the input field  in Eq.~(\ref{eq:langevin}) is linear. Moreover, for the purpose of calculating $ \overline{ \hat{c}_{o,j} } $, we can replace the operators $\hat{c}^{\rm in}_{\nu,j}$ in the source terms of the  Langevin equation (\ref{eq:langevin}) with their mean values
$\overline{\hat{c}^{\rm in}_{\nu,j}}$. We can even formally replace all the input terms with the coherent interaction
\be\label{Hpert}
H_I=i\sqrt{\kappa_{o,k}}\left(\hat{c}^\dagger_{o,k}\alpha_p\exp[-i \Omega t]-\hat{c}_{o,k}\alpha^*_p\exp[i \Omega t]\right).
\ee
Thus,  $ \overline{ \hat{c}_{o,j} } $ is  given by the Kubo formula where $H_I$ plays the role of the perturbation. We note that the optomechanical coupling  does not conserve the number of excitations
and, therefore, $ \overline{ \hat{c}_{o,j} } $ has both an elastic (frequency $ \Omega $ in the rotating frame or
$ \Omega + \omega_L$ in the laboratory frame) and an inelastic (frequency $ - \Omega $ in the rotating frame or
$-\Omega + \omega_L$ in the laboratory frame) components. The elastic part of transmission is obtained
after time averaging:
\begin{equation}
   T_{oo}(j,k;\Omega) = \left|
          \frac{\Omega}{2\pi} \int_0^{\frac{2\pi}{\Omega}}
                     \frac{ \overline{ \hat{c}^{\rm out}_{o,j} }(t) }{ \overline{ \hat{c}^{\rm in}_{o,k} }(t) } {\rm d} t
                   \right|^2 .
\end{equation}
Using Eqs.(\ref{NormAmp}--\ref{Hpert}) and the Kubo formula, we  find
\begin{equation}
\label{Too}
   T_{oo}(j,k;\Omega)=\left|\delta_{j,k}-i \, \sqrt{\kappa_{o,j}\kappa_{o,k}} \, G^R_{oo}(j,k;\Omega)\right|^2 \, ; \quad
   G_{oo}(j,k;\Omega)=-i \int_0^\infty {\rm d} t \, e^{i \Omega t} \, \overline{ [\hat{c}_{o,j}(t),\hat{c}^\dagger_{o,k}(0)] } .
\end{equation}
For $j \neq k$, we recover the formula for $ T_{oo}(j,k;\Omega) $ given in the main text.

If the eigenstates of the OMA are localized then $ T_{oo} (j,k; \Omega) $ decays exponentially
on large distances and the inverse localization length can defined as follows:
\be
  \label{LocRad-QM}
  \frac{1}{\xi(\Omega)} = - \lim_{ x \to \infty} \frac{ \ln \bigl[ T_{oo} (j,k; \Omega) \bigr] }{ 2 x } =
                              - \lim_{ x \to \infty} \frac{ \ln | G^R_{oo} (j,k; \Omega ) | }{ x } ; \quad
                               x \equiv | j - k |.
\ee
We note that Eq.(\ref{LocRad-QM}) contains only matrix elements of the Green's function relating operators
from sites $j$ and $k$. These elements can be calculated iteratively with the help of the Dyson's
equation which is similar to that suggested in Ref.[S2] for the transfer matrix (details of the algorithm can
be found in Ref.[S3]).

\begin{figure}[t]
   \includegraphics[width=0.7 \textwidth]{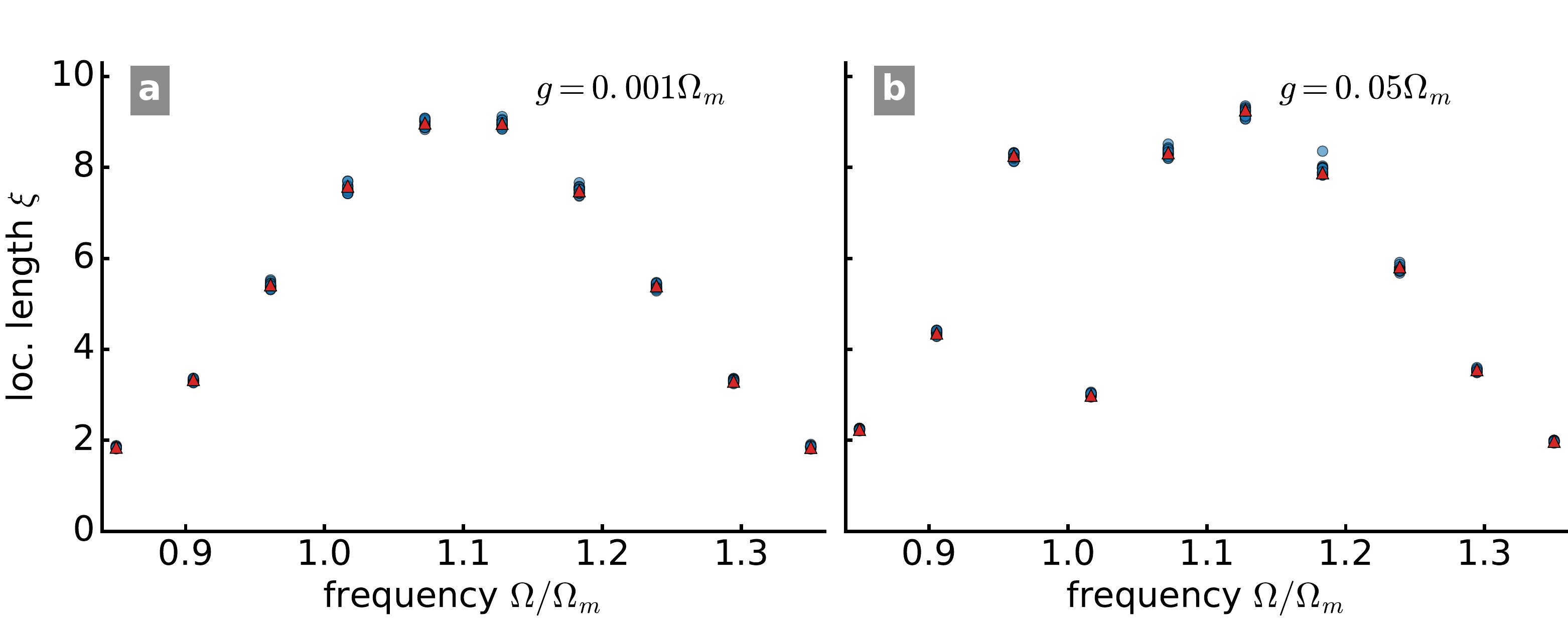}
   \caption{\label{16-loc-rad}
                  Blue dots: 16 localization lengths obtained after inserting each entry of $ \hat{{\cal G}}^R $ into
                  Eq.(\ref{LocRad-QM}). Red dots are obtained from Eq.(\ref{LocLenForNum}).
                  Ten different values of the frequency have been taken to demonstrate that
                  all components of the matrix Green's function show the same spatial decay.
                  Parameters of the OMA are the same as in the main text: $ \Omega_o = 1.1 \Omega_m,
                  J_o = \sigma_o = 0.1 \Omega_m, J_m = \sigma_m = 0.001 \Omega_m $,  $ g = 0.001 \Omega_m $
                  (left panel) and $ g = 0.05 \Omega_m $ (right panel). Numerics were done in the absence of dissipation.
                 }
\end{figure}

Generically, one can introduce 4 transmissions in the elastic channel, $ T_{\nu\nu'} (j,k; \Omega) $, and 4 transmissions
in the inelastic one, $ \tilde{T}_{\nu\nu'} (j,k; \Omega) $ (e.g., the photon-photon transmission, inelastic photon-photon
transmission). Similar to Eq.(\ref{Too}), these transmissions are described by entries of the matrix Green's function
constructed from four-component "Opto-mechanical$\times$Nambu"-spinors $ \hat{C}_{j} $:
\bea
  \label{4-Spinors}
    \hat{C}_{j}^{\rm T}(t) & = & \left\{ \hat{c}_{o,j}(t), \hat{c}_{m,j}(t), \hat{c}_{o,j}^\dagger(t), \hat{c}_{m,j}^\dagger(t) \right\}; \\
  \label{MatrixGF}
    \hat{{\cal G}}^R(j, t; j', t') & = & \ri \theta(t-t')
           \overline{
              \hat{C}_{j}(t) \otimes \hat{C}_{j'}^\dagger(t') -
              \hat{C}_{j'}^\ast(t') \otimes \hat{C}_{j}^T(t)
                    } .
\eea
$ \hat{{\cal G}}^R $ has 16 entries which can be obtained from a straightforward generalization of the Dyson
equation for the matrix Green's function in frequency space, $ \hat{{\cal G}}^R (j, j'; \Omega) $. We have solved the generalized
Dyson equation numerically for the parameters of the OMA given in the main text and compared 16 largest localization
lengths governed by inserting each component of $ \hat{{\cal G}}^R $ into Eq.(\ref{LocRad-QM}). These localization lengths
coincide up to small numerical errors $ \le 1\% $, see blue dots in Fig.\ref{16-loc-rad}.
This is the related to the symmetries, namely, particle-antiparticle symmetry, time-reversal symmetry,
and space-inversion symmetry. The latter appears effectively in the long disordered OMAs due to the 
self-averaging. The equivalence of the different transmissions on large distances allows one to find 
the largest localization lengths of the OMA from any convenient linear combination of 
$ | \hat{{\cal G}}^R_{ab} (j, j'; \Omega) | $ ensuring a good convergence of the numerical algorithm. In particular, 
we can use ``the generalized transmission'' of the opto-mechanical excitations
\be
\label{Prop-Tr}
    {\cal T}(j,k;\Omega) = {\rm Tr} \left( | \hat{{\cal G}}^R (j, k; \Omega) |^2 \right) ;
\ee
and, after disorder averaging, arrive at:
\be
\label{LocLenForNum}
    \frac{1}{\xi_2(\Omega)} = - \lim_{ x \to \infty} \frac{ \la \ln \bigl[ {\cal T}(j,k; \Omega ) \bigr] \ra_d }{ 2 x } , \quad
    x \equiv | j - k | .
\ee
Eq.(\ref{LocLenForNum}) has been used in the numerical code with the disorder averaging being
substituted by the self-averaging of $ \xi $ in very long systems, see red dots in Fig.\ref{16-loc-rad}.

\subsubsection*{2.2 Bogoluibov eigenstates}

In the absence of dissipation, there is a simple method which allows
one to find the localization length directly from the average number of the excitation.
This approach is realized after diagonalizing the Hamiltonian (or, equally, the Heisenberg
equations of motion) with the help of the Bogoliubov transformation. Let us define
eigenmode operators
$\hat{d}_s$.
Generically, $\hat{d}_s$ can be written as follows:
\begin{equation}
  \hat{d}_s = \sum\limits_{j=1}^{N} \sum_\nu \left[ u_{j,s}^{(\nu)} \, \hat{c}_{\nu,j} +
                                                           v_{j,s}^{(\nu)} \, \hat{c}_{\nu,j}^{\dagger} \right], \quad 1 \leq s \leq 2 N \, .
  \label{a1}
\end{equation}
Here $ u_{j,s}^{(\nu)} $ and $ v_{j,s}^{(\nu)} $ are the Bogoliubov coefficients. The
transformation matrix that diagonalize the Heisenberg equations reads as:
\begin{equation}
  T=\left[\begin{array}{ll}
  U^{(o)} & \left[ V^{(o)} \right]^\ast   \\
  U^{(m)} & \left[ V^{(m)} \right]^\ast \\
  V^{(o)} & \left[ U^{(o)} \right]^\ast   \\
  V^{(m)} & \left[ U^{(m)} \right]^\ast
  \end{array}\right],
  \label{a2}
\end{equation}
where $U^{(\nu)}$ and $V^{(\nu)}$ are $N \times 2 N$ matrices whose entries are the coefficients $ u_{j,s}^{(\nu)} $
and $ v_{j,s}^{(\nu)} $ from Eq.(\ref{a1}).
In the absence of dissipation and instabilities, these coefficients satisfy the following relation:
\begin{equation}
 \sum\limits_{j=1}^{N} \sum_\nu  \left\{ u_{j,s}^{(\nu)} \left[ u_{j,s'}^{(\nu)} \right]^* -
                                                               v_{j,s}^{(\nu)} \left[ v_{j,s'}^{(\nu)} \right]^* \right\} = \delta_{s, s'}.
 \label{a4}
\end{equation}
The minus sign in front of summands $  v_{j,s}^{(\nu)} \left[ v_{j,s'}^{(\nu)} \right]^* $ is caused by the bosonic commutation
relations of the operators $ \hat{c}_{\nu,j} $. As a consequence, $T$ is pseudounitary with the inverse matrix
\begin{equation}
  T^{-1}=\left[\begin{array}{cccc}
   \left[ U^{(o)} \right]^\dagger & \left[ U^{(m)} \right]^\dagger  & -\left[ V^{(o)} \right]^\dagger & -\left[ V^{(m)} \right]^\dagger \\
  -\left[ V^{(o)} \right]^{\rm T} & -\left[ V^{(m)} \right]^{\rm T} &  \left[ U^{(o)} \right]^{\rm T} &  \left[ U^{(m)} \right]^{\rm T}
  \end{array}\right] .
  \label{a5}
\end{equation}
The time evolution of the operators $\hat{c}_{\nu,j}$ can be obtained using Eq. (\ref{a5}), and it is given by
\begin{equation}
\label{c-exp}
  \hat{c}_{\nu,j}(t) = \sum\limits_{s=1}^{2 N}    \left\{
    \left[ u_{j,s}^{(\nu)} \right]^\ast \hat{d}_s(0) e^{-i \varepsilon_s t} - v_{j,s}^{(\nu)} \hat{d}_s^{\dagger}(0) e^{i \varepsilon_s t}
                                                \right\}.
\end{equation}
Here $ \varepsilon_s $ is the frequency of the  hybrid (opto-mechanical) eigenmode $ s $. Using Eqs.(\ref{c-exp}),
one can derive
\be
  \langle 0| \hat{d}_s \hat{n}_{\nu,j} \hat{d}_s^{\dagger} |0 \rangle =
         \left| u_{j,s}^{(\nu)} \right|^2 + \left| v_{j,s}^{(\nu)} \right|^2 + \sum_s \left| v^{(\nu)}_{j,s} \right|^2 ; \quad
  \langle 0| \hat{n}_{\nu,j} |0 \rangle = \sum_s \left| v^{(\nu)}_{j,s} \right|^2 ;
\ee
and find the total average number of excitations at a given site $j$ after the eigenmode $s$ 
is excited:
\begin{equation}
  n_j(s) =
  \langle 0| \hat{d}_s \hat{n}_j \hat{d}_s^{\dagger} |0 \rangle - \langle 0| \hat{n}_j |0 \rangle =
                        \sum_\nu \left( \left| u_{j,s}^{(\nu)} \right|^2 + \left| v_{j,s}^{(\nu)} \right|^2 \right) , \
  \hat{n}_j = \hat{n}_{o,j} + \hat{n}_{m,j}.
  \label{a7.1}
\end{equation}
We have subtracted the (background) fluctuations of $ \hat{n}_j $ in the ground state since the number of
excitation in the OMA always fluctuates due to the optomechanical coupling, see the second term in the RHS
of Eq.(\ref{eq:cellHamiltonian}).

The eigenmodes of the OMA can be found via the numerical diagonalization of the Heisenberg equations.
Now we substitute $ n_j(s) $ for $ {\cal T}_{oo}(j,k,\Omega) $ in Eq.(\ref{LocLen}) and associate the frequency $ \Omega $
with $ \varepsilon_s $ and the origin $ k $ with the coordinate where the eigenmode $ s $ has maximal amplitude. This yields
the second expression for $ \xi_2 $. Thus, the localization length can be estimated from a log-linear fit of $ n_j(s) $,
see the discussion of Eq.(\ref{LocLen}) in the main text.

\subsection*{3. Weak coupling regime, $ \Delta_{\rm loc}^{\rm (m)} \leq g \leq J_m \ll \sqrt{J_m J_o} $}

\begin{figure}[ht]
   \includegraphics[width=0.45 \textwidth]{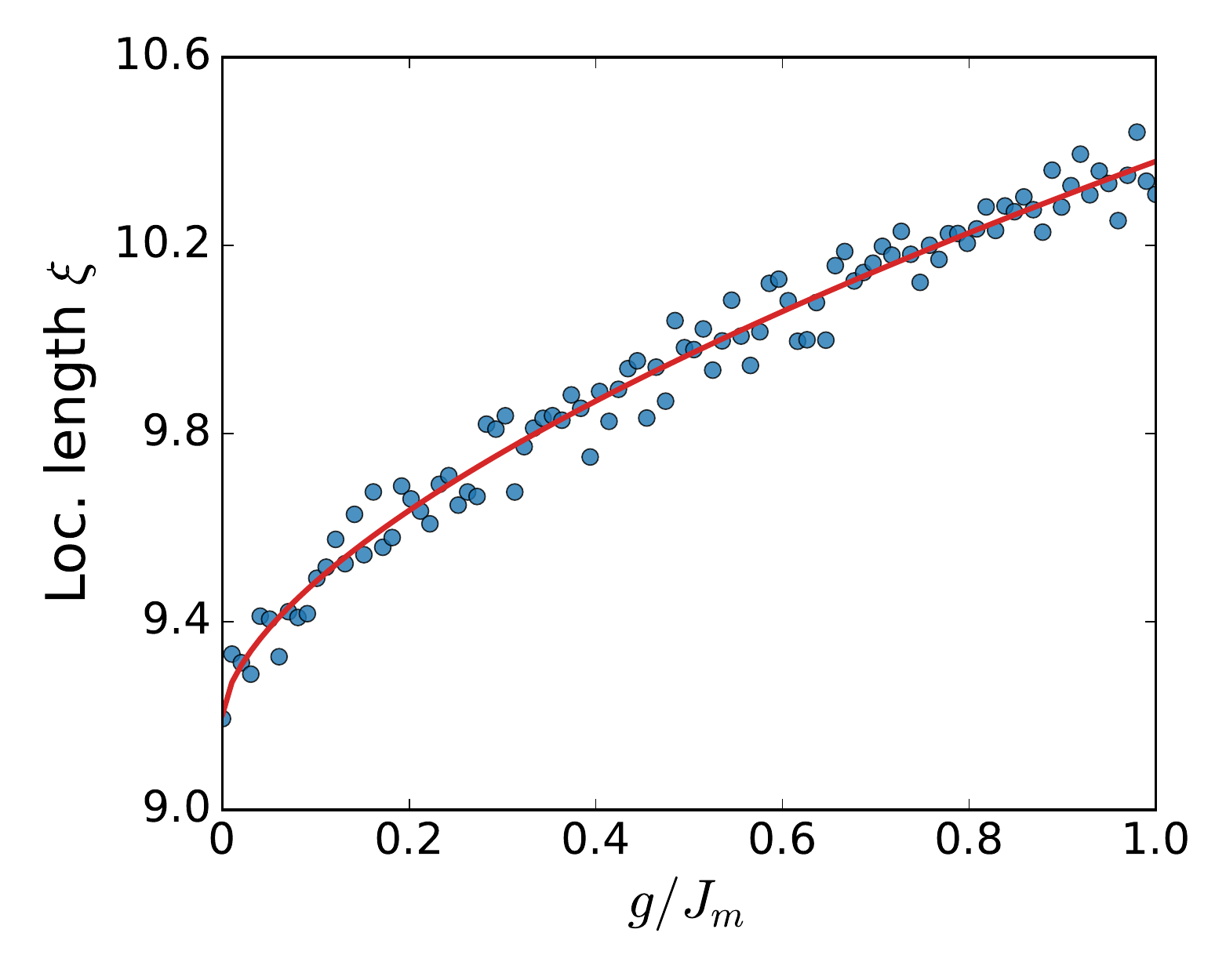}
   \qquad
   \includegraphics[width=0.4 \textwidth]{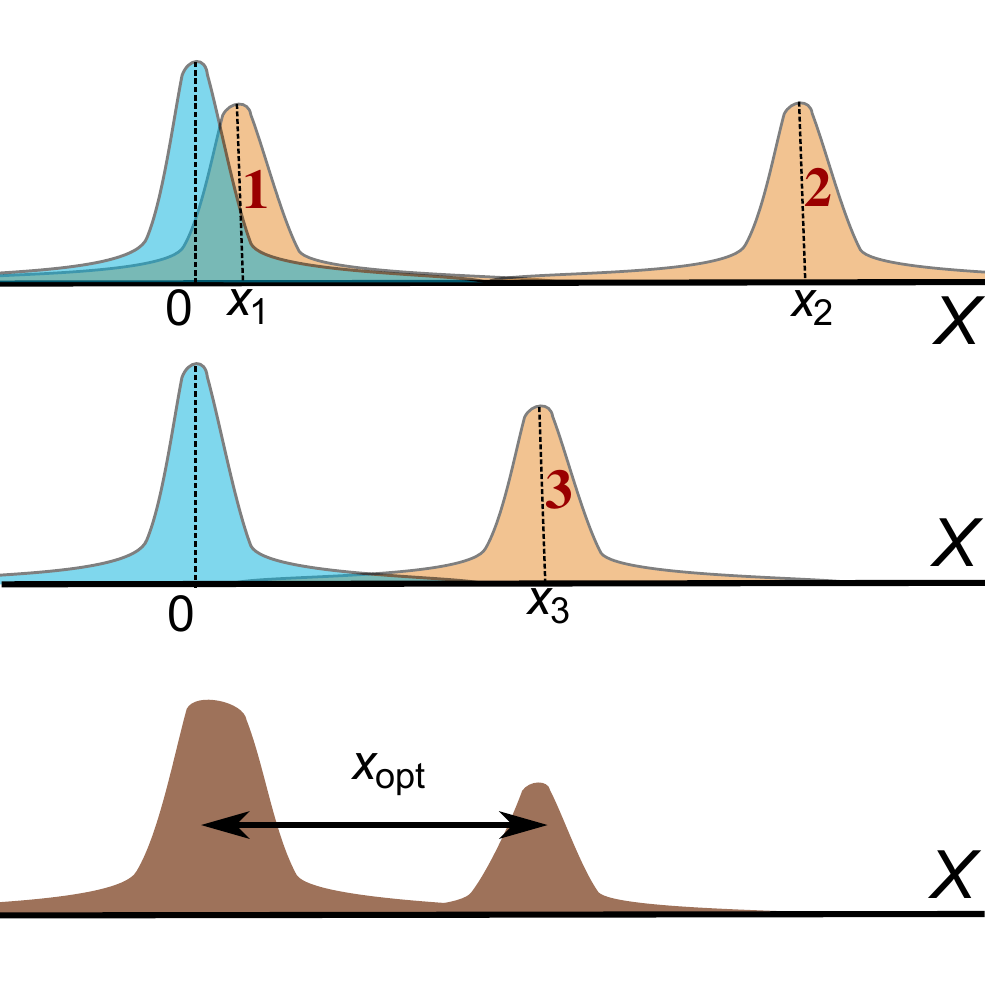}
   \caption{\label{Xi-g}
                  Left panel: Numerically obtained dependence $ \xi_2(g) $ for $ g \le J_m, \, \Omega=\Omega_m $ (shadowed dots).
                  The red line is an example of the fitting which demonstrates the sub-linear nature of this dependence.
                  Right panel: a bare optical state (with blue filling)
                  can be hybridized with different bare mechanical states (with orange filling). The hybridization
                  with the state No.1 is strong but it is unable to change the largest localization length substantially.
                  The hybridization with the state No.2 is negligible. The hybridization with the (optimal) state 
                  No.3 is also strong and is responsible for the increase of $ \xi_2 $. The optical state being hybridized
                  with the mechanical states No.1,3 yields a ``double-hump'' optomechanical state (with brown filling)
                  which is responsible for the transmission $ {\cal T}_{om} $ on large distances.
                 }
\end{figure}

Let us analyze the behavior of $ \xi_2(g) $ for the case $ g \leq J_m \ll \sqrt{J_m J_o} $ where
the influence of the optomechanical coupling on the band structure is negligible. The numerical
analysis shows that $ \xi_2(g) $ is sub-linear, see the left panel of Fig.\ref{Xi-g}, which indicates
the presence of non-perturbative contributions.
The full theory for this is missing and we give only phenomenological arguments which are
similar to those of the Mott theory [S4] and allow one to explain the sub-linear growth of $ \xi_2 $
when $ g $ increases up to $ J_m $. For simplicity, we concentrate on the transmission 
$ {\cal T}_{om} $ in the regime $ \Delta_{\rm loc}^{\rm (m)} \sim J_m / \xi_m^{(0)} \lesssim g \ll J_m $. 
Other parameters correspond to Fig.\ref{LocLenEn}a in the main text.

Finite transmission $ {\cal T}_{om} $ requires hybridization of bare optical and mechanical states.
The main idea of the phenomenological approach is to find a pair of the optical- and the mechanical-
states which, being strongly hybridized, provides the largest possible increase of $ \xi_2 $. In other 
words, we have to estimate the maximal distance between bare localization centers which does not 
violate the necessary condition for the strong hybridization.

Consider an optical state with the frequency inside the unperturbed  mechanical band, $ \Omega_m - J_m 
< \epsilon_o < \Omega_m + J_m $, see the blue wave-function in the right panel of Fig.\ref{Xi-g}. 
The space coordinates will be counted from the localization center of this optical state. Such a 
state can be strongly hybridized with the mechanical states if inequality
\be
\label{ResCond}
  g \, \la M_j | O \ra \ge | \epsilon_o - \epsilon_{m,j} | ,
\ee
holds true. Here $ j $ is the number of the mechanical state with the localization center at $ x_j > 0 $
and with the frequency $ \epsilon_{m,j} $; 
$ \la M_j | O \ra $ is the overlap between the localized optical and the localized mechanical states
\be
  \la M_j | O \ra \sim \left[ \xi_{m}^{(0)} \exp\left( - x_j /\xi_{m}^{(0)} \right)  - 
                                           \xi_{o}^{(0)} \exp\left( - x_j /\xi_{o}^{(0)} \right) \right] \Bigl/
                                         \left( \xi_{m}^{(0)} - \xi_{o}^{(0)} \right) .
\ee
We recall that $ \xi_{m}^{(0)} > \xi_{o}^{(0)} $ for $ \Omega \simeq \epsilon_o $, cf. Fig.\ref{LocLenEn}a.

Firstly we note, that, unlike the Mott theory, frequencies $ \epsilon_o $ and $ \epsilon_{m,j} $ are not 
correlated at $ g = 0 $. Therefore, $  | \epsilon_o - \epsilon_{m,j} | $ can be arbitrary small even if 
the localization centers of the bare states are close to each other, $ x_j \ll \xi_{o,m}^{(0)} \Rightarrow 
\la M_j | O \ra \sim 1 $, see the orange state No.1 in the right panel of Fig.\ref{Xi-g}. On the other 
hand, it is clear that the 1st mechanical state is unable to support an essential increase of the 
transmission $ {\cal T}_{om} $ beyond the bare localization length.

$ {\cal T}_{om} $ can become more long-ranged if $ \xi^{(0)}_m \lesssim x_j $. In the
extreme case $ \xi_{o}^{(0)} \ll \xi_{m}^{(0)} \ll  x_j $, the overlap becomes exponentially small, 
$ \la M_j | O \ra \sim \exp( - x_j / \xi^{(0)}_m) $. Distant mechanical states do not obey the condition 
Eq.(\ref{ResCond}) and, therefore, are unimportant, cf. the orange state No.2 in the right panel of 
Fig.\ref{Xi-g}. However, there is always an optimal state for which $ x_j $ is relatively large and the 
smallness of $ \la M_j | O \ra $ in Eq.(\ref{ResCond}) is compensated by the smallness of the frequency 
separation:
\be
\mbox{optimal state:} \quad
  \la M_j | O \ra \sim \frac{ | \epsilon_o - \epsilon_{m,j} | }{ g } 
         \ \Rightarrow \
  x_{\rm opt} \sim \xi_m^{(0)} \log\left( g \Bigl/ \Delta^{\rm (m)}_{\rm loc} \right) ;
\ee
cf. the orange state No.3 in the right panel of Fig.\ref{Xi-g}. Now we can speculate that, if $ \Delta_{\rm loc}^{\rm (m)}
\sim J_m / \xi_m^{(0)} \lesssim g \ll J_m $, $ {\cal T}_{om} $ on large distances and, correspondingly, $ \xi_2 $ are 
governed by the ``double-hump'' optomechanical state originating mainly from hybridization of the optical state with 
the optimal mechanical one, see an example in the right panel of Fig.\ref{Xi-g}. Therefore, the largest localization length 
can be estimated as
\be
  \label{Xi-g-Log}
  \xi_2 \simeq \xi_m^{(0)} + {\rm const} \times x_{\rm opt} \simeq 
      \xi_m^{(0)} \left[ C_ 1 +  C_2 \log\left( g \Bigl/ \Delta^{\rm (m)}_{\rm loc} \right) \right] \, .
\ee
($ C_{1,2} $ are constants of order $ O(1) $ which cannot be determined in the frame of the phenomenological approach).
The optomechnical state may be  ``multiple-hump'' if $ x_3 $ covers several localization volumes 
of the bare mechanical states. If $ g \ll \Delta_{\rm loc}^{\rm (m)} $ we expect a crossover to the 
purely perturbative regime.

Since $ \Delta^{\rm (o)}_{\rm loc} \gg J_m \sim \sigma_m $, the minimal space distance between two
optical states belonging to the frequency range of the mechanical band is large. We estimate it as $ \xi_o^{(0)}
\Delta^{\rm (o)}_{\rm loc} / J_m \gg  x_{\rm opt} $. This condition allows us to consider relevant
optical states independently. 

The universal dependence Eq.(\ref{Xi-g-Log}) can be justified only if the frequency range from
$ \Delta^{(m)}_{\rm loc} $ to $ J_m $ is broad and $ x_{\rm opt} \gg \xi_m^{(0)} $. This is not 
the case for the parameters of the main text, in particular, because the disorder is strong.
The fitting in the left panel of Fig.\ref{Xi-g} can be done equally by using 
either Eq.(\ref{Xi-g-Log}) or a power-law dependence with some non-universal exponent 
$ \alpha < 1 $. More rigorous theory of the weak coupling regime can be developed 
by exploiting basic ideas of the virial expansion, see Refs.\cite{YeKr-virial} and [S5,S6] for 
more details.


\subsection*{4. Localization of hybrid excitations in the case of fluctuating coupling constant}

\begin{figure}[ht]
   \includegraphics[width=0.38 \textwidth]{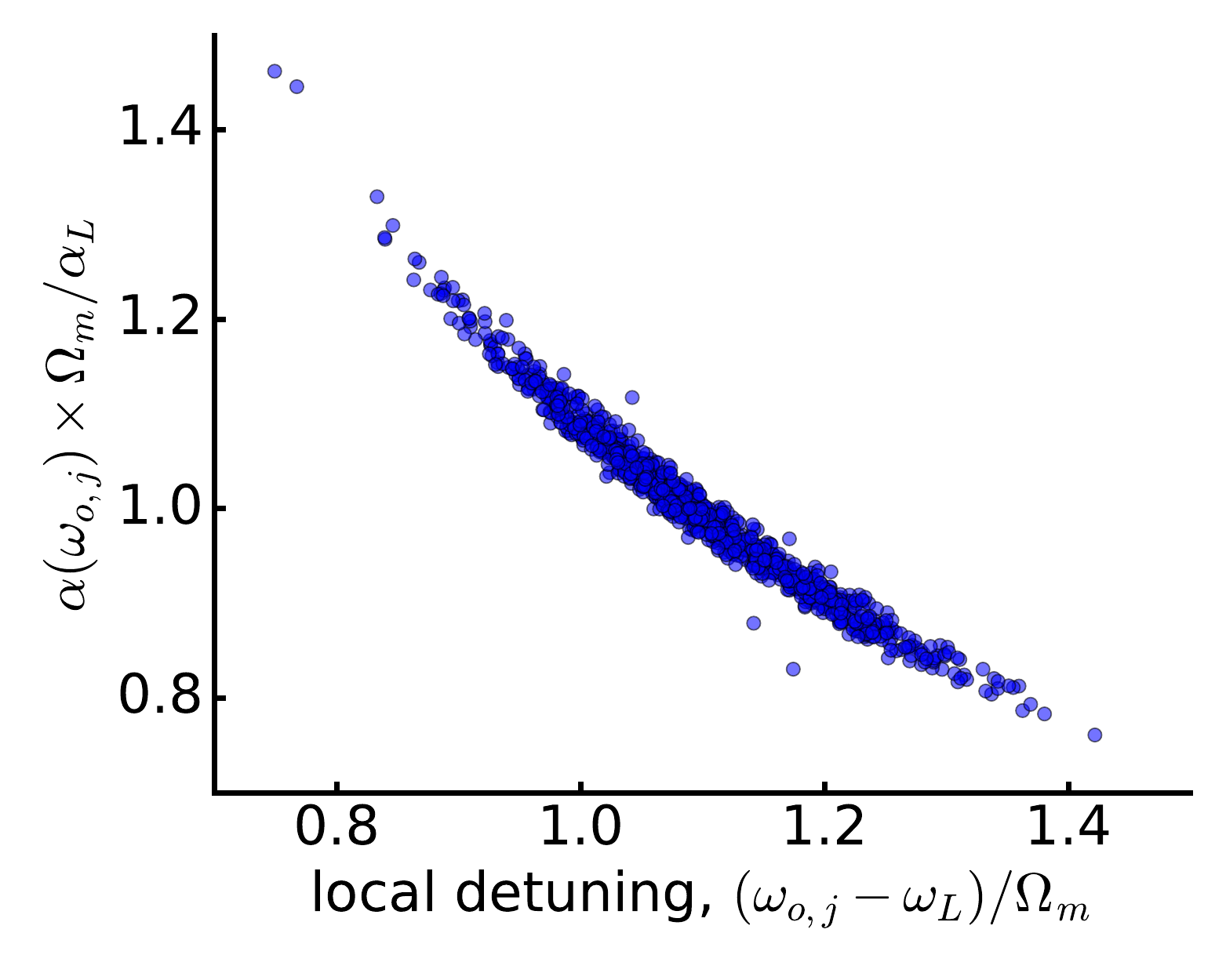}
   \caption{\label{AlphaEn}
                  Dependence of $ \alpha(\omega_{0,j}) = c_{o,j}|_{g_0=0}$ on the optical frequency of the cells.
                 }
\end{figure}

In the main text, we have concentrated on the case where the optomechanical coupling
$ g $ is one and the same for all cells. In reality, the coupling fluctuates: $ g_j $
depends on the mean occupation number of the photons on the site $ j $, $ g_j \propto c_{o,j} $,
while the cells with smaller optical frequencies host more photons, see Fig.\ref{AlphaEn}. This
locally enhances $ g_j $ on these sites. One can speculate that the coupling constant acquires an
effective frequency dependence; $ g $ becomes larger for smaller frequencies and it slightly decreases
with increasing the frequency. We note that $ \alpha(\omega_{0,j}) $ depicted in Fig.\ref{AlphaEn}
is defined as $ \alpha(\omega_{0,j}) = c_{o,j}|_{g_0=0}$.
We do not
distinguish $ \alpha(\omega_{0,j}) $ and $ c_{o,j}$ since their difference is small, $  (\alpha(\omega_{0,j}) - c_{o,j} )
\sim O(g_0^2) $.

\begin{figure}[ht]
   \includegraphics[width=0.7 \textwidth]{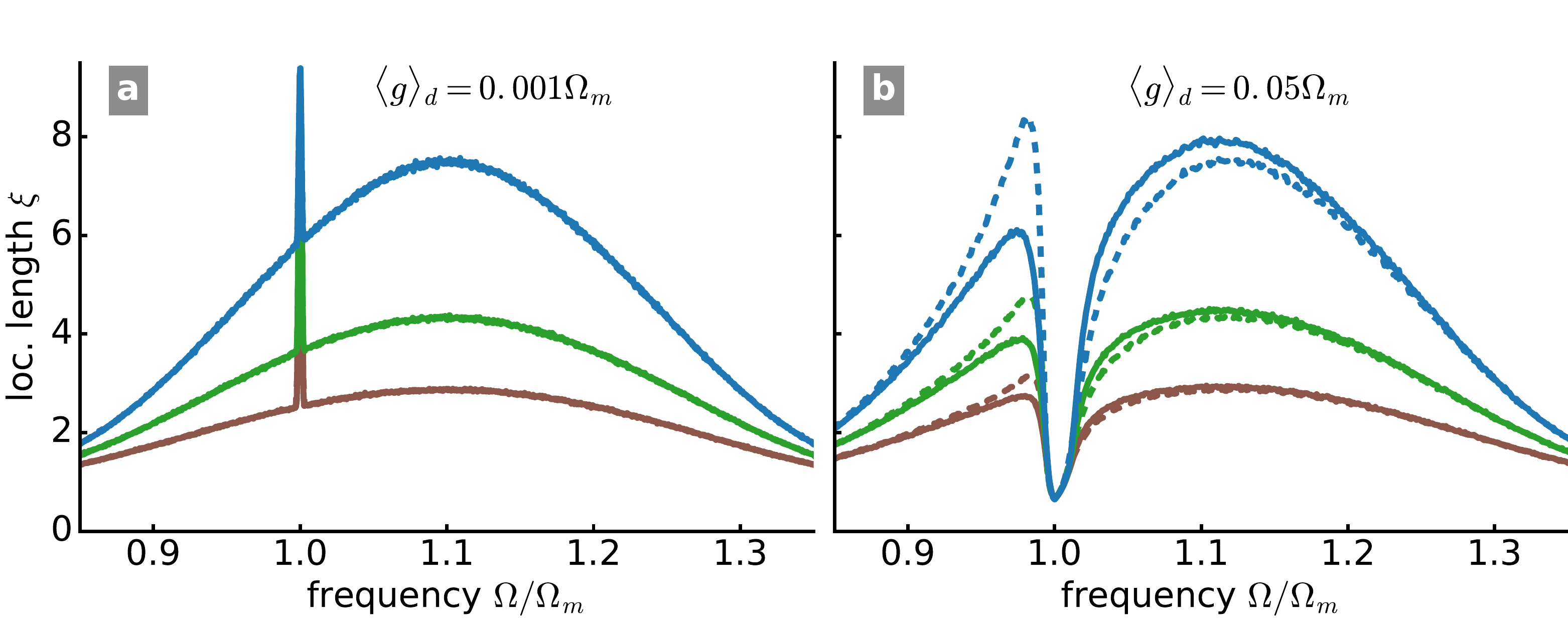}
   \caption{\label{Asymm}
        Frequency dependence of the localization length at $ \Omega_o = 1.1 \Omega_m $
        and $ \la g \ra_d = g = \{ 0.001, 0.05 \} \Omega_m $ 
        calculated for different values of the optical decay rate
        $ \kappa_o / \Omega_m = \{ 0.01, 0.05, 0.1 \} $ (blue-, green-, and brown lines,
        respectively). Solid/dashed lines show profiles with/without fluctuations of the
        coupling constant.
                 }
\end{figure}

We have recalculated curves from Fig.\ref{Dissip} for fluctuating  $ g_j $. Results
are shown in Fig.\ref{Asymm}. The disorder averaged coupling has been adjusted
to the same values (a) $ \, \la g \ra_d = g = 0.001 \Omega_m $;  and (b) $ \la g \ra_d
= g = 0.05 \Omega_m $. Solid/dashed lines show the frequency
dependence of the localization length with/without fluctuations of the coupling.
The profiles at smaller $ g $ are almost intact by its fluctuations.
When the mean coupling is larger, the left (right) maximum of $ \xi_2(\Omega) $ is suppressed (enhanced)
by these fluctuations. This can be explained if, based on Fig.\ref{AlphaEn}, we assume
that $ g $ becomes effectively larger (smaller) at $ \Omega < \Omega $ ($ \Omega > \Omega $) and notice
that the peaks decrease when $ g $ increases, cf. Fig.\ref{LocLenEn}.
Thus, the fluctuations of $ g_j $ are able to modify the shape of $ \xi_2(\Omega) $ at relatively large values
of the mean coupling constant $ g $ though all qualitatively important features are expected to be robust.

\newpage

\begin{center}
  \hrule
\end{center}

\vspace{0.25 cm}

\begin{enumerate}
   \item[ [S1\!\!] ] C. C. Gerry, P. L. Knight, {\it Introductory quantum optics},
                    Cambridge Univ. Press (2005).
   \item[ [S2\!\!] ] A. MacKinnon and B. Kramer, \prl \ {\bf 47}, p.1546 (1981).
   \item[ [S3\!\!] ] A. MacKinnon and B. Kramer, Zeitschrift f{\"u}r Physik B
                          Condensed Matter {\bf 53},  pp.1-13 (1983).
   \item[ [S4\!\!] ] N. F. Mott, Philosophical Magazine {\bf 22} pp.7-29 (1970).
   \item[ [S5\!\!] ] O. Yevtushenko, and A. Ossipov, J. Phys. A: Math. Theor. {\bf 40}, pp.4691-4716  (2007).
   \item[ [S6\!\!] ] S. Kronm{\"u}ller, O.M. Yevtushenko, and E. Cuevas, J. Phys. A: Math. Theor. {\bf 43}, 075001 (2010).
\end{enumerate}


\end{document}